\documentclass[aps,prd,showpacs,epsfig]{revtex4}
\usepackage[dvips]{graphicx}

\begin{document}

\title{Feedback effects on the pairing interaction in color superconductors 
near the transition temperature}

\author{Kei Iida}
\affiliation{RIKEN BNL Research Center, Brookhaven National Laboratory,
Upton, New York 11973, USA}
\date{\today}

\begin{abstract}
     We examine the role that the gap dependence of the pairing interaction
plays in the gap equation for a weakly coupled uniform superfluid of 
three-flavor massless quarks near the transition temperature $T_c$.  We find 
that the feedback effects on Landau-damped transverse gluons mediating the 
pairing interaction alter the gap magnitude in a way dependent on the
color structure of the gap.  We estimate corrections by these effects to 
the parameters characterizing the fourth order terms in the Ginzburg-Landau 
free energy and ensure the stability of a color-flavor locked state near 
$T_c$.

\end{abstract}


\maketitle

\section{Introduction}
\label{sec:intro}

      It has been noted since the seminal work in Refs.\ \cite{barrois,BL} 
that the quark-quark interaction in the color antitriplet channel is attractive
and drives a Cooper pairing instability in quark matter in the limit of high 
density in which the Fermi energy of the quarks dominates over the one-gluon 
exchange interaction energy.  Because of the relativistic nature of 
high density quark matter, the color magnetic (transverse) force plays a 
dominant role in Cooper pairing.  This interaction is nonlocal
in time just like the electron-phonon interaction in ordinary superconductors
\cite{nambu,eliashberg}, but in contrast is 
long-ranged in the absence of static screening.
This long-range nature gives rise to a different dependence of the pairing gap
on the QCD coupling constant $g$ from the BCS result \cite{son}.  At zero 
temperature, up to leading order in $g$, the logarithm of the gap arises 
from the dynamically screened magnetic force that involves Landau damped 
virtual gluons in a normal medium.  Corrections to the gluon self-energy 
(polarization function) by the gap do not affect the logarithm of the gap
up to subleading order in $g$ at zero temperature \cite{rischke}.  This is 
because the gluon self-energy is modified by the gap significantly only at 
gluon energies of $\lesssim T_c$.  The influence of such corrections on the 
gap equation near the transition temperature $T_c$, however, has yet to be 
examined.

      In this paper we investigate the structure of the gap equation near 
$T_c$ by including the polarization effects of the color superconducting 
medium on exchanged gluons.  We then estimate corrections thereby induced to
the parameters characterizing the fourth order terms in the Ginzburg-Landau 
free energy of a weakly coupled uniform superfluid of massless three-flavor 
quarks.  These corrections, divided by the weak coupling value, are of 
order $g$, in contrast with the case of a short-range pairing interaction in 
which the corrections generally contain a factor proportional to the ratio of 
the transition temperature to the Fermi energy.  We find that the polarization 
corrections keep the color-flavor locked phase the most stable just below 
$T_c$.  Throughout this paper, we consider a system of three-flavor ($uds$) 
and three-color ($RGB$) massless quarks at temperature $T$ and baryon chemical
potential $\mu$, and use units $\hbar=c=1$.  We assume that the Fermi momentum
is common to all colors and flavors.

\section{Gap equation}
\label{sec:gap}

     In this section we address the question of how the gap equation 
relevant in the weak coupling regime, in which the pairing interaction 
is induced by one-gluon exchange, is modified by the polarization effects
of the color superconducting medium.  For this purpose we first consider a 
$J^P=0^+$ pairing state that is $ud$-isoscalar and $RG$-color antitriplet, 
since this is one of the simplest states that belong to a color and flavor 
antisymmetric channel with $J^P=0^+$.  This channel has a common transition 
temperature in the limit of weak coupling \cite{BLR}:
\begin{equation}
  T_c= \frac{2e^\gamma}{\pi} e^{-(\pi^2+4)/8}\  ~\frac{b\mu}{3} 
  ~{\rm exp} \left(-\frac{3\pi^2}{\sqrt{2}  g}\right),
  \label{tc}
\end{equation}
where ${e^\gamma}/{\pi} = 0.5669\ldots$ and $b=256 \pi^4 (2/3 g^2)^{5/2}$.
For the isoscalar pairing state of interest here, a nonzero excitation gap 
$d({\bf k})$ is open below $T_c$ for quasiparticles having either flavor 
$u$ or $d$ and either color $R$ or $G$.  Here ${\bf k}$ is the momentum 
associated with the relative coordinate of a quark Cooper pair.  In the case
in which only the modification of the one-gluon exchange force by a normal 
medium is included in the random-phase approximation (RPA) \cite{weldon,BMPR} 
and the normal state Hartree-Fock correction to the quark propagator is 
ignored, the gap equation reads \cite{PR,I}
\begin{eqnarray}
 d({\bf k})
 &=&\frac{g^{2}}{48\pi^{3}}
 \int d^3 q
[D_T(E({\bf q})-E({\bf k}),{\bf q}-{\bf k})
+D_T(E({\bf q})+E({\bf k}),{\bf q}-{\bf k})
 \nonumber \\ & & 
+D_L(E({\bf q})-E({\bf k}),{\bf q}-{\bf k})
+D_L(E({\bf q})+E({\bf k}),{\bf q}-{\bf k})]
 \nonumber \\ & &
 \times
 d({\bf q})
 E^{-1}({\bf q})\tanh\left(\frac{E({\bf q})}{2T}\right)\ ,
  \label{deq}
\end{eqnarray}
where
\begin{equation}
   E({\bf q})= \left[\left(|{\bf q}|-\frac{\mu}{3}\right)^{2}+
   d^2({\bf q})\right]^{1/2}
  \label{Eofq}
\end{equation}
is the excitation energy, 
\begin{equation}
  D_T(p)\simeq
  {\rm Re}\left[
     \frac{1}
  {|{\bf p}|^{2}-i\pi m_{D}^{2}p_{0}\theta(\sqrt{\pi}m_{D}/2-|{\bf p}|)
   /4|{\bf p}|}\right]
  \label{gluePRT}
\end{equation}
and
\begin{equation}
  D_L(p)\simeq
     \frac{1}{|{\bf p}|^{2}+m_{D}^{2}},
  \label{gluePRL}
\end{equation}
with the Debye screening mass
\begin{equation}
   m_{D}=\left[\frac{3g^{2}}{2\pi^{2}}\left(\frac{\mu}{3}\right)^2
     +\frac{3g^{2}T^{2}}{2}\right]^{1/2},
 \label{debye}
\end{equation}
characterize the transverse and longitudinal parts of the gluon propagator
$D(p)$ in the Landau gauge as 
\begin{equation}
D^{\alpha\beta}_{\mu\nu}(p)=-\delta_{\alpha\beta}
[P^T_{\mu\nu}D_T(p)+P^L_{\mu\nu}D_L(p)],
\end{equation}
with the transverse and longitudinal projection operators
\begin{equation}
  P^{T}_{ij}=\delta_{ij}-\frac{p_{i}p_{j}} {|{\bf p}|^{2}}, ~~
  P^{T}_{00}=P^{T}_{0i}=P^{T}_{i0}=0,
\end{equation}
\begin{equation}
    P^{L}_{\mu\nu}=\frac{p_{\mu}p_{\nu}}{p^{2}}-g_{\mu\nu}-P^{T}_{\mu\nu}.
\end{equation}
Expressions (\ref{gluePRT}) and (\ref{gluePRL}) are approximate in the
sense that they are available in the regime $p_0\ll|{\bf p}|\ll\mu/3$, 
but duly allow for the Landau damping of transverse virtual gluons and
the Debye screening of the longitudinal force in a way sufficient to describe 
the exact form of the logarithm of the gap magnitude up to subleading order in
$g$ \cite{PR}.  Note that the Landau damping provides an effective infrared 
cutoff in the transverse sector, $\sim(\pi m_D^2 |p_0|/4)^{1/3}$, which in 
turn plays a dominant role in determining the pairing gap.

     At $T=0$, the solution to the gap equation (\ref{deq}) is known as 
\cite{PR}
\begin{equation}
    d({\bf k})=\frac23 b\mu e^{-\pi/2{\bar g}} \sin({\bar g} x),
  \label{prgap}
\end{equation}
where ${\bar g}\equiv g/3\sqrt2 \pi$, and
\begin{equation}
   x\equiv\ln\left[\frac{2b\mu/3}{||{\bf k}|-\mu/3|+E({\bf k})}\right].
  \label{x}
\end{equation}
The factor $\sin({\bar g}x)$ ensures that the gap is appreciable only for
momenta ${\bf k}$ close to the Fermi surface.  The exponential term and 
the sinusoidal $x$ dependence arise from nearly static, Landau-damped 
magnetic gluons that mediate the long-range part of the magnetic interactions,
while both the higher frequency magnetic gluons and Debye-screened electric 
gluons play a dominant role in determining the pre-exponential factor.

     For later comparison with the case near $T_c$, we write down the 
equation for the magnitude of the gap on the Fermi surface,
$d_F\equiv d(|{\bf k}|=\mu/3)$, which can be derived from Eq.\ (\ref{deq})
as \cite{PR}
\begin{equation}
  d_F=\frac{2g^2}{(3\pi^2)^2}
      \left[\ln^2\left(\frac{2\delta}{d_F}\right)
           +b'\ln\left(\frac{2\delta}{d_F}\right)\right]d_F.
  \label{gapeq0}
\end{equation}
Here the cutoff $\delta$, obeying $d_F\ll\delta \ll m_D$, is chosen so that 
$d(|{\bf k}|>\delta)$ is vanishingly small, and $b'=2\ln(b\mu/3\delta)$.
The term associated with $\ln^2(2\delta/d_F)$ comes from soft Landau-damped 
magnetic gluons, while the term associated with $\ln(2\delta/d_F)$ comes from 
nonstatic magnetic gluons and Debye-screened electric gluons.

     The overall coefficient of the $\mu/g^5$ in the pre-exponential factor 
in Eq.\ (\ref{prgap}) is correct up to a factor of order unity since 
the quasiparticle wave function renormalization ignored here results in a 
factor $\exp[-(\pi^2+4)/8]$ \cite{wang}, which appears also in the weak 
coupling expression for $T_c$, Eq.\ (\ref{tc}).  (This renormalization affects
the sinusoidal $x$ dependence only through a factor of order $g^2 x$.)  On the 
other hand, polarization by the color superconducting medium, which gives 
rise to the gap dependence of the pairing interaction, provides even higher 
order corrections to the zero-temperature gap \cite{rischke}.

     Near $T_c$, the momentum dependence of the gap can be set equal to 
that at $T=0$, as in the usual BCS case \cite{PR}.  Consequently,
\begin{equation}
  d({\bf q},T)=d_F(T)\sin({\bar g}y),
\end{equation}
where 
\begin{equation}
   y\equiv\ln\left[\frac{2b\mu/3}{||{\bf q}|-\mu/3|+E({\bf q},T=0)}\right].
  \label{y}
\end{equation}
Then, the gap magnitude $d_F$ can be determined by expanding the gap equation 
(\ref{deq}) up to ${\cal O}(d^3)$ as
\begin{eqnarray}
  d_F&=&\frac{g^{2}}{18\pi^{2}}
  \int_{0}^{\delta} d(|{\bf q}|-\mu/3)
  \ln\left(\frac{b\mu/3}{||{\bf q}|-\mu/3|}\right)
 \nonumber \\ &&
 \times \left\{
   d({\bf q})
 \frac{\tanh\left(||{\bf q}|-\mu/3|/2T\right)}{||{\bf q}|-\mu/3|} 
  +d^3({\bf q})
 \frac{1}{2||{\bf q}|-\mu/3|}\frac{d}{d||{\bf q}|-\mu/3|}
 \left[\frac{\tanh\left(||{\bf q}|-\mu/3|/2T\right)}{||{\bf q}|-\mu/3|}
   \right]+\cdots
   \right\},
\end{eqnarray}
where we have noted that the momentum region, ${\bf k}\approx{\bf q}$, 
contributes dominantly to the integral in Eq.\ (\ref{deq}).  We thus obtain 
\begin{equation}
   d_F =  \left(1-\frac{\pi{\bar g}}{2}\frac{T-T_c}{T_c}\right)d_F
    -\frac{7\zeta(3){\bar g}}{16 \pi T_c^2} d_F^3 +  {\cal O}(d_F^5),
  \label{geqnct}
\end{equation}
with the zeta function $\zeta(3)=1.2020\ldots$.  Here the coefficients 
affixed to $d_F$ and $d_F^3$ include the leading contributions
with respect to $g$, and we have noted that at $T=T_c$, 
\begin{eqnarray}
  1&=&\frac{g^{2}}{18\pi^{2}}
  \int_{0}^{\delta} d(|{\bf q}|-\mu/3)
  \ln\left(\frac{b\mu/3}{||{\bf q}|-\mu/3|}\right)
   \sin({\bar g}y)
 \frac{\tanh\left(||{\bf q}|-\mu/3|/2T_c\right)}{||{\bf q}|-\mu/3|}
  \label{thouless}
\end{eqnarray}
is satisfied.  [In the absence of the quasiparticle wave function
renormalization, the solution to Eq.\ (\ref{thouless}) reproduces 
expression (\ref{tc}) except for a factor $\exp[-(\pi^2+4)/8]$.]
The solution to Eq.\ (\ref{geqnct}) reads 
\begin{equation}
    d_F=\left[\frac{8\pi^2 T_c^2}{7\zeta(3)}\frac{T-T_c}{T_c}\right]^{1/2}.
  \label{sol0}
\end{equation}
As it should, Eq.\ (\ref{geqnct}) is the same as the known result obtained
from the Ginzburg-Landau theory (see the next section).

     We now introduce the effect of the color superconducting medium on the 
gluon propagator within the RPA.  The normal gluon propagator characterized
by Eqs.\ (\ref{gluePRT}) and (\ref{gluePRL}) is modified by the pairing gap,
in a way dependent on $\alpha$, as \cite{dirk}
\begin{equation}
  D_T^{\alpha}(p)\simeq
  {\rm Re}\left[
     \frac{1}
  {|{\bf p}|^{2}+(m_M^\alpha)^2 f({\bf p})
  -i\pi m_{D}^{2}p_{0}\theta(\sqrt{\pi}m_{D}/2-|{\bf p}|)
   /4|{\bf p}|}\right]
  \label{gluePRT2}
\end{equation}
and
\begin{equation}
  D_L^{\alpha}(p)\simeq
     \frac{1}{|{\bf p}|^{2}+m_{D}^{2}-3(m_M^\alpha)^2 h({\bf p})}.
  \label{gluePRL2}
\end{equation}
Here only leading corrections by the pairing gap have been retained, and
\begin{equation}   
 (m_M^\alpha)^2 = \left\{
 \begin{array}{ll}
  0, & \quad \mbox{$\alpha=1,2,3,$} \\
  g^2 K_T d_F^2, & \quad \mbox{$\alpha=4,5,6,7,$} \\
  (4/3)g^2 K_T d_F^2, & \quad \mbox{$\alpha=8,$}
 \end{array}
\right.
   \label{mei2SC}
\end{equation}
are the Meissner screening masses \cite{II} with the stiffness parameter in the
weak coupling limit:
\begin{equation}
  K_T=\frac{7\zeta(3)}{24(\pi T_c)^2}N\left(\frac{\mu}{3}\right),
    \label{KT}
\end{equation}
where 
\begin{equation}
   N\left(\frac{\mu}{3}\right)=\frac{1}{2\pi^2}\left(\frac{\mu}{3}\right)^2
    \label{DOS}
\end{equation}
is the ideal gas density of states at the Fermi surface.  $f({\bf p})$ and 
$h({\bf p})$ are the dimensionless positive definite functions that
characterize the ${\cal O}(d^2)$ corrections to the transverse and 
longitudinal parts of the irreducible particle-hole bubble \cite{dirk}.  These 
functions reduce to unity in the limit of ${\bf p}\to0$, while decreasing to 
zero with increasing $|{\bf p}|$.  We note that the Landau damping term,
corresponding to the energy-dependent term in Eq.\ (\ref{gluePRT2}), does
undergo corrections by a factor of $1+{\cal O}(d^2/|{\bf p}|^2)$, but they 
lead to higher order corrections to the gap equation in $T_c/\mu$ 
as compared with those coming from the Meissner term 
$(m_M^{\alpha})^2 f({\bf p})$.

     Expressions (\ref{gluePRT2}) and (\ref{gluePRL2}) are the straightforward
extension of the normal medium forms (\ref{gluePRT}) and (\ref{gluePRL})
to the case of the color superconducting medium.  These 
expressions for $D_T$ and $D_L$ retain consistency with the transverse and 
longitudinal sum rules obeyed by the static, long-wavelength gluon propagator 
in normal quark matter \cite{PR} and in color superconducting quark matter 
\cite{II}.

     It is important to note that the leading feedback effect of the pairing
gap lies in the magnetic sector.  This is partly because the corrections to 
the gluon propagator are roughly of order $m_M^2/(m_D^2 |p_0|)^{2/3}$ in the 
magnetic sector, while of order $m_M^2/m_D^2$ in the electric sector, and 
partly because the gluon energy range $|p_0|\ll m_D$ plays a dominant role 
in the gap equation.  In deriving such corrections in the magnetic sector, 
we first expand $D_T$, Eq.\ (\ref{gluePRT2}), with respect to $m_M^2$, and 
then substitute into $f({\bf p})$ the form relevant near $T_c$:
\begin{equation}
  f({\bf p})=\frac{6}{7\zeta(3)}\sum_{s=0}^{\infty}
     \int_0^1 dx \frac{1-x^2}{(s+1/2)[4(s+1/2)^2+(|{\bf p}|x/2\pi T_c)^2]}.
    \label{f}
\end{equation}
This form is identical to that encountered in the usual BCS case \cite{LL} 
since in both cases a pairing gap is open for quasiparticle momenta so close 
to the Fermi surface that quasiparticles and quasiholes having momenta 
${\bf k}_1$ and ${\bf k}_2$ with ${\bf k}_1+{\bf k}_2={\bf p}$ and 
$|{\bf k}_1|\simeq|{\bf k}_2|\simeq \mu/3$ dominate the ${\cal O}(d^2)$ 
corrections to the particle-hole bubble.  In the London limit 
($|{\bf p}|\to0$), one can set $f\simeq1$, while in the Pippard limit 
($|{\bf p}|\to\infty$) $f$ behaves as $\propto 1/|{\bf p}|$.  
As we shall see, the Pippard regime ($|{\bf p}|>2\pi T_c$) is as important to
the gap equation as the London regime ($|{\bf p}|<2\pi T_c$).

     The gap equation modified by the leading feedback effect can be obtained 
by replacing $D_T$ by
\begin{equation}
  D_T^{\alpha}(p)\simeq
  {\rm Re}\left[
     \frac{1}
  {|{\bf p}|^{2}-i\pi m_{D}^{2}p_{0}\theta(\sqrt{\pi}m_{D}/2-|{\bf p}|)
   /4|{\bf p}|}\right]
   -(m_M^\alpha)^2 f({\bf p})
   {\rm Re}\left\{
     \frac{1}
  {[|{\bf p}|^{2}-i\pi m_{D}^{2}p_{0}\theta(\sqrt{\pi}m_{D}/2-|{\bf p}|)
   /4|{\bf p}|]^2}\right\}
  \label{gluePRT3}
\end{equation}
in Eq.\ (\ref{deq}).  The modification associated with $m_M^\alpha$ works 
only for $\alpha=8$.  This is partly because the Meissner masses 
vanish for $\alpha=1$--3 [see Eq.\ (\ref{mei2SC})] and partly because the 
contributions of gluons of $\alpha=4$--7 to the gap equation vanish due to 
the color structure of the gap \cite{PR}.  Near $T_c$, the gap equation at 
$|{\bf k}|=\mu/3$ thus reads
\begin{eqnarray}
  d_F&=&\left(1-\frac{\pi{\bar g}}{2}\frac{T-T_c}{T_c}\right)d_F
    -\frac{7\zeta(3){\bar g}}{16 \pi T_c^2} d_F^3 
 \nonumber \\ &&
 +\frac{g^{2}}{8\pi^{2}}
  \int_{0}^{\delta} d(|{\bf q}|-\mu/3)
   d({\bf q})
 \frac{\tanh\left(||{\bf q}|-\mu/3|/2T_c\right)}{||{\bf q}|-\mu/3|} 
   \left(\frac{1}{12}\right) (m_M^{\alpha=8})^2 
    \frac{1}{2|{\bf q}|^2}F({\bf q}),
 \nonumber \\ &&
+  {\cal O}(d_F^5),
\label{geqnctmod}
\end{eqnarray}
where
\begin{equation}
F({\bf q})=\frac13 \int_{-1}^{1}d(\cos\theta)
    \frac{1-\cos\theta}{(1-\cos\theta)^3
   + (\pi m_D^2 ||{\bf q}|-\mu/3|/8\sqrt2 |{\bf q}|^3)^2
     \theta(\sqrt{\pi}m_{D}/2-\sqrt2|{\bf q}|\sqrt{1-\cos\theta})}
      \left[f(y)+y\frac{df(y)}{dy}\right]
  \label{fq}
\end{equation}
with $y=|{\bf q}|\sqrt{1-\cos\theta}/\sqrt2 \pi T_c$.  Here,
$\theta$ is the angle between ${\bf k}$ and ${\bf q}$, we have replaced 
$|{\bf k}|$ by $\mu/3$ in the gluon propagator, and the factor $1/12$ 
comes from the color vertex part of $\alpha=8$.  Note that the leading
term by the feedback effects in Eq.\ (\ref{geqnctmod}) is of third order
in $d_F$ and hence does not affect $T_c$.

     We proceed to examine the leading corrections to the gap equation due to 
the color superconducting medium.  For this purpose, it is useful to divide 
the gluon momentum $|{\bf p}|\simeq\sqrt2 |{\bf q}|\sqrt{1-\cos\theta}$ and 
energy $p_0\simeq||{\bf q}|-\mu/3|$ into several regimes shown in Fig.\ 1.
The boundary $y=1$ corresponds to the gluon momentum $|{\bf p}|=2\pi T_c$.
For $y<1$ (London regime) and $y>1$ (Pippard regime), $f(y)$ has 
the following forms:
\begin{equation}
  f(y)=\sum_{n=0}^{\infty} a_n y^{2n}, ~~a_n=\frac{24}{7\zeta(3)}
   \frac{(-1)^n \zeta(2n+3)}{(2n+1)(2n+3)}\left[1-2^{-(2n+3)}\right],
   \label{fy1}      
\end{equation}
and
\begin{equation}
  f(y)=\frac{6}{7\zeta(3)y}\sum_{s=0}^{\infty}\left\{
    \left[\frac{1}{2(s+1/2)^2}+\frac{2}{y^2}\right]\tan^{-1}\frac{y}{2(s+1/2)}
   -\frac{1}{(s+1/2)y}\right\}.
   \label{fy2}      
\end{equation}
At $y=1$, as can be seen from Eq.\ (\ref{fq}), the effective infrared 
cutoff due to the Landau damping of transverse gluons is nonnegligible when 
$p_0$ is larger than $32\pi^2 T_c^3/m_D^2$.  Another important scales are 
$p_0=\pi T_c$, above which $\tanh(p_0/2T_c)$ in Eq.\ (\ref{geqnctmod})
approaches unity exponentially, and $|{\bf p}|=\sqrt\pi m_D/2$, above which 
transverse gluons no longer undergo Landau damping.  
We note that the gap can be regarded as flat in regions a) and c).
We also note that the timelike regime ($p_0>|{\bf p}|$), in which the gluon
propagator is not described well by the form (\ref{gluePRT3}), can be safely 
ignored since in this regime no feedback corrections to the gap equation occur 
up to leading order in $T_c/\mu$.  In fact, the gap corrections to 
the gluon propagator vanish like $\sim d^2/p_0^2$ with increasing $p_0$,
as in the $T=0$ case \cite{rischke}.

\begin{figure}[t]
\begin{center}
\includegraphics[width=10cm]{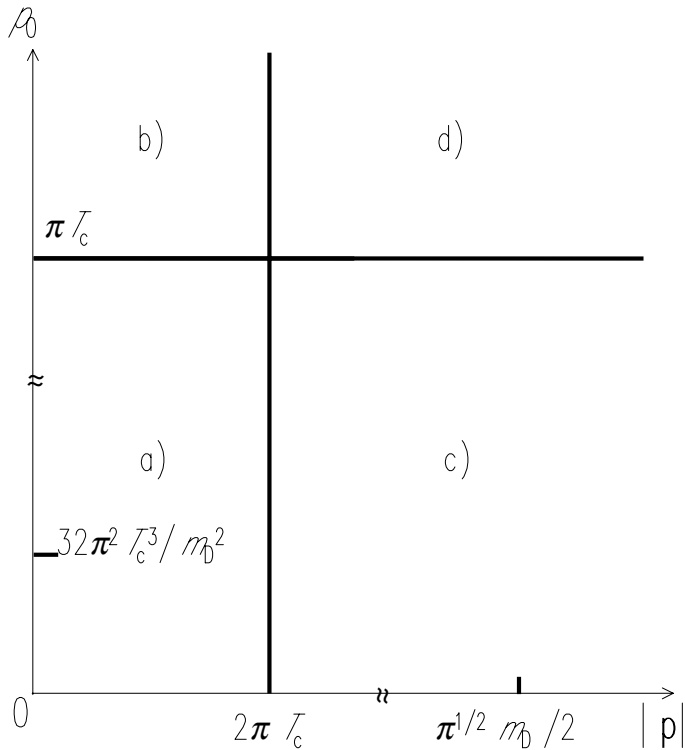}
\end{center}
\vspace{-0.5cm}
\caption{\label{fig1}
The energy-momentum regimes of exchanged magnetic gluons. 
}
\end{figure}

     At $2\pi T_c < |{\bf p}|$, we can use expression (\ref{fy2}) for $f(y)$.
From this expression, we obtain
\begin{equation}
 f(y)+y\frac{df(y)}{dy} =\frac{12}{7\zeta(3)y^2} 
   {\cal F}(y),
   \label{fydfy0}      
\end{equation}
with
\begin{equation}
 {\cal F}(y)=
  \sum_{s=0}^{\infty}\left[-\frac{2}{y}\tan^{-1}\frac{y}{2(s+1/2)}
   +\frac{1}{s+1/2}\right].
\end{equation}
The large $y$ asymptotic behavior of $f(y)+ydf(y)/dy$ is then
\begin{equation}
  f(y)+y\frac{df(y)}{dy}\approx\frac{12}{7\zeta(3)y^2}\ln y.
    \label{fydfy}
\end{equation}
This behavior is different from $\propto y^{-1}$, which is followed by
$f(y)$.  As a result, transverse gluons of momenta near $T_c$ rather than 
near the Pippard limit are essential to calculations of the feedback effect 
in the magnetic sector.  This is a contrast to the case of the weak coupling 
limit in which transverse gluons of momenta large compared with $T_c$ dominate
the pairing interaction since for such momenta, the factor 
$m_D^2 p_0/|{\bf p}|$ characterizing the Landau damping in the
propagator (\ref{gluePRT}) is sufficiently small that the
pairing interaction remain essentially long ranged.

     Using Eqs.\ (\ref{fq}) and (\ref{fy2}), we calculate the contribution
to the gap equation (\ref{geqnctmod}) from regions c) and d).  The momentum
range covering these regions corresponds to the range of $\theta$ satisfying 
$-1 < \cos\theta < 1-2\pi^2 T_c^2/|{\bf q}|^2$.  Up to leading order in $g$,
the result from region c) reads 
\begin{equation}
  \frac{\pi[3\pi^3-28\zeta(3)f(1)]{\bar g}^2 (m_M^{\alpha=8})^2 d_F}
       {112\zeta(3)m_D^2},
   \label{regionc}
\end{equation}
while the contribution from region d) is of higher order in $T_c/\mu$ and 
thus can be ignored.

     At $|{\bf p}|<2\pi T_c$, where expression (\ref{fy1})
is available for $f(y)$,  $f(y)$ and hence $f(y)+ydf(y)/dy$ are almost
flat.  In $F({\bf q})$, this momentum range corresponds to the range of 
$\theta$ satisfying $1-2\pi^2 T_c^2/|{\bf q}|^2 < \cos\theta < 1$.  
The contribution to the gap equation (\ref{geqnctmod}) from region a)
becomes, to leading order in $g$,
\begin{equation}
 \frac{\pi{\bar g}^2 (m_M^{\alpha=8})^2 d_F}{4m_D^2}\sum_{n=0}^\infty a_n,
  \label{regiona}
\end{equation}
while that from region b) is of higher order in $T_c/\mu$.
It is remarkable that the term (\ref{regiona}) is comparable to the 
term (\ref{regionc}).  This suggests that transverse gluons of momenta 
below and above $2\pi T_c$ are equally important to the feedback effect.  
We also note that these terms are of order $g^2 d_F^3/T_c^2$ and thus
suppressed only by one power $g$ with respect to the term proportional 
$d_F^3$ in the gap equation (\ref{geqnct}) in the weak coupling limit.
This is a contrast to the case of a short-range pairing force in which the 
leading correction to the third order term due to the superfluid medium is 
suppressed by one power $T_c/\mu$ \cite{He3}.  
We remark that the scale of the gluon energy dominant in the gap equation
is of order $32\pi^2T_c^3/m_D^2$.  Since this is much smaller than the typical
momentum scale $\sim 2\pi T_c$, we can safely take the static limit
of the modification due to the color superconducting medium.

    We finally rewrite the gap equation (\ref{geqnctmod}) in such a way that 
the feedback term has a coefficient up to leading order in $g$, i.e., by
combining the contributions (\ref{regionc}) and (\ref{regiona}) from regions 
c) and a).  The result is 
\begin{equation}
  d_F=\left(1-\frac{\pi{\bar g}}{2}\frac{T-T_c}{T_c}\right)d_F
    -\frac{7\zeta(3){\bar g}}{16 \pi T_c^2} d_F^3 
    +\frac{7\zeta(3)C{\bar g}^2}{32\pi T_c^2} d_F^3 
    +  {\cal O}(d_F^5),
\label{geqnctmod2}
\end{equation}
where $C=\pi^3/63\zeta(3)=0.409434\ldots$.  We thus find that the leading 
feedback effect near $T_c$ acts to increase the gap squared of the isoscalar 
pairing state by a factor of $(1-C{\bar g}/2)^{-1}$.  This is due to the fact 
that the feedback effect manifests itself as Meissner screening of the color 
magnetic force of color index $\alpha=8$; this force is repulsive in 
contrast to the attractive case of $\alpha=1$--3 dominating the pairing 
interaction.  We remark that in the $T=0$ case in which expansion of the gap 
equation with respect to the gap magnitude is not valid, the terms associated 
with the logarithm of the gap in the gap equation (\ref{gapeq0}) mainly 
determine the gap magnitude.  In this case, the feedback effect provides 
corrections beyond these logarithmic terms \cite{rischke}.

     For the purpose of calculating corrections to the Ginzburg-Landau
parameters in the next section, it is instructive to repeat the above
calculations for a color-flavor locked (CFL) state, one of the $J^P=0^+$, 
color and flavor antisymmetric pairing states.  In the CFL state, all three 
flavors and colors are equally gapped in such a way that the pairing gap 
between a quark of color $a$ and flavor $i$ and a quark of color $b$ and 
flavor $j$ is characterized by 
$\kappa(\delta_{ai}\delta_{bj}-\delta_{aj}\delta_{bi})$.  For the on-shell 
gap on the Fermi surface, $\kappa_F$, the gap equation near $T_c$ can be 
written in the form similar to Eq.\ (\ref{geqnctmod2}) as
\begin{equation}
  \kappa_F=\left(1-\frac{\pi{\bar g}}{2}\frac{T-T_c}{T_c}\right)\kappa_F
    -\frac{7\zeta(3){\bar g}}{8 \pi T_c^2} \kappa_F^3 
    -\frac{21\zeta(3)C{\bar g}^2}{8\pi T_c^2} \kappa_F^3 
    +  {\cal O}(\kappa_F^5).
\label{geqcfl}
\end{equation}
Here we have used the Meissner masses in the CFL state \cite{II}, i.e., 
$(m_M^\alpha)^2=2g^2 K_T \kappa_F^2$ for $\alpha=1$--8.  In the limit
of $m_M^\alpha\to0$, Eq.\ (\ref{geqcfl}) is equivalent to Eq.\ (104) in 
Ref.\ \cite{I}.  We thus find that the leading feedback effect acts to reduce
the gap squared by a factor of $(1+3C{\bar g})^{-1}$.  This reduction stems
from the fact that Meissner screening of the color magnetic force
takes effect equally for $\alpha=1$--8.  Note a contrast with the case
of the isoscalar pairing state in which the feedback effect acts to 
increase the gap magnitude.

\section{Ginzburg-Landau free energy}
\label{sec:GL}

     We proceed to derive the Ginzburg-Landau free energy of a weakly coupled 
uniform superfluid of massless three-flavor quarks from the gap equation near 
$T_c$ as examined in the previous section.  Instead of focusing on the 
isoscalar and CFL pairing states, it is convenient to construct the 
Ginzburg-Landau free energy for a more general color and flavor antisymmetric 
channel with $J^P=0^+$ as in Ref.\ \cite{I}.  This is because all states 
belonging to this channel has a common value of $T_c$, which reduces to Eq.\ 
(\ref{tc}) in the weak coupling limit.  This channel is characterized by 
a complex $3\times3$ gap matrix, $({\bf d}_a)_i$, in color-flavor 
space \cite{I}, where $a$ ($i$) is the color (flavor) other than two
colors (flavors) involved in Cooper pairing.  This gap is defined on the mass 
shell of the quark quasiparticle of momenta on the Fermi surface, and thus 
reduces to $\delta_{aB}\delta_{is}d_F$ in the isoscalar state and to
$\delta_{ai}\kappa_F$ in the CFL state.  For $({\bf d}_a)_i$, one can write 
down the thermodynamic potential density difference 
$\Delta\Omega=\Omega_s-\Omega_n$ between the superfluid and normal phases 
near $T_c$ as \cite{I}
\begin{equation}
   \Delta\Omega= \bar{\alpha} \sum_{a}|{\mathbf{d}}_a|^2
   +\beta_1(\sum_{a}|{\mathbf{d}}_a|^2)^2
   +\beta_2 \sum_{ab}|{\mathbf{d}}_a^{\ast}\cdot {\mathbf{d}}_b|^2.
   \label{gl}
\end{equation}
Here each term is invariant with respect to $U(1)$ global gauge 
transformations and color and flavor rotations.

     In evaluating the coefficients in Eq.\ (\ref{gl}) by including the 
leading feedback effect, it is convenient to integrate the gap equations 
(\ref{geqnctmod2}) and (\ref{geqcfl}) and then map the results onto Eq.\ 
(\ref{gl}).  We thus obtain
\begin{eqnarray}
 {\bar\alpha}&=&4N(\mu/3)\ln\left(\frac{T}{T_{c}}\right),  \\
 \beta_{1}&=&\frac{7\zeta(3)}{8(\pi T_c)^{2}}
          \left(1+\frac{13}{2}C{\bar g}\right)
   N(\mu/3), 
 \label{beta1n} \\
 \beta_2&=&\frac{7\zeta(3)}{8(\pi T_c)^{2}}
          \left(1-\frac{15}{2}C{\bar g}\right)
   N(\mu/3),
 \label{beta2n}
\end{eqnarray}
which reproduce the known relation $\beta_1=\beta_2$ in the weak coupling 
limit \cite{I}.  We find that up to leading order in $g$, the polarization 
effects of the color superconducting medium give rise to only ${\cal O}(g)$ 
corrections of the coefficients $\beta_1$ and $\beta_2$.  It is nonetheless
important to note that these corrections to $\beta_1$ and $\beta_2$ work in 
the opposite directions in such a way as to decrease and increase the gap
magnitude in the CFL and isoscalar states, respectively.  We can thus 
conclude that whether or not the leading feedback effect acts to reduce the 
gap magnitude depends on the color structure of the gap.

\section{Phase diagram}
\label{sec:phase}

     We turn to the construction of the phase diagram in the space of
the parameters characterizing the fourth order terms in the Ginzburg-Landau
free energy derived in the previous section.  This phase diagram can be 
obtained by minimizing the thermodynamic potential difference $\Delta\Omega$ 
with respect to $({\bf d}_a)_i$ for various values of $\beta_1$ and $\beta_2$
\cite{I}.  The result, exhibited in Fig.\ 2, is the same as Fig.\ 1 in Ref.\ 
\cite{I} except that the present values of $\beta_1$ and $\beta_2$ given by 
Eqs.\ (\ref{beta1n}) and (\ref{beta2n}) include the ${\cal O}(g)$ feedback 
corrections.  This figure shows that even for such values of $\beta_1$ and 
$\beta_2$, as long as ${\bar g}$ is sufficiently smaller than unity, the 
CFL phase, which generally satisfies
\begin{equation}
 {\bf d}_{R}^{*}\cdot{\bf d}_{G}={\bf d}_{G}^{*}\cdot{\bf d}_{B}
     ={\bf d}_{B}^{*}\cdot{\bf d}_{R}=0, ~~
 |{\bf d}_{R}|^{2}=|{\bf d}_{G}|^{2}=|{\bf d}_{B}|^{2},
 \label{opcfl}
\end{equation}
is still favored over the two-flavor color superconducting (2SC) state 
fulfilling
\begin{equation}
  {\bf d}_{R}\parallel{\bf d}_{G}\parallel{\bf d}_{B}.
 \label{opis}
\end{equation}
Note that the 2SC state contains the isoscalar state analyzed in Sec.\ 
\ref{sec:gap}.

\begin{figure}[t]
\begin{center}
\includegraphics[width=15cm]{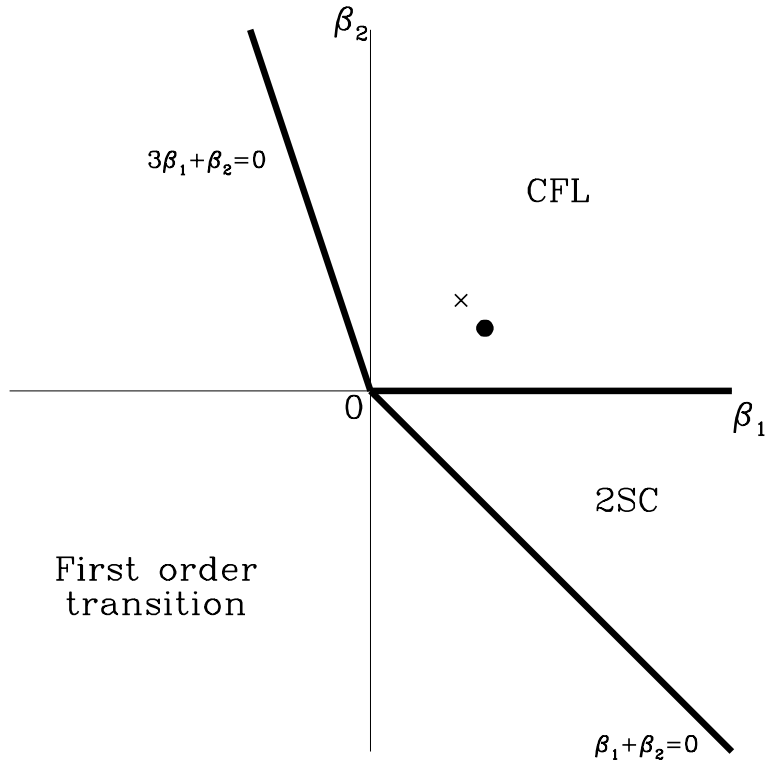}
\end{center}
\vspace{-0.5cm}
\caption{\label{fig2}
Phase diagram near $T_c$, exhibiting regions where the 2SC and CFL 
phases are reached by a second order transition as well as where 
superfluid states are reached by a first order transition since the
overall fourth order term in the Ginzburg-Landau free energy (\ref{gl}) can 
be negative.  The parameters $\beta_1$ and $\beta_2$ are the fourth order 
coefficients in the Ginzburg-Landau free energy.  The cross denotes 
the weak coupling limit, and the circle denotes the result including
the polarization effects of the color superconducting medium with 
${\bar g}=0.1$.
}
\end{figure}

\section{Conclusions}
\label{sec:concl}

     We have examined the role played by the gap dependence of the pairing 
interaction in the gap equation for a weakly coupled uniform superfluid of 
three-flavor massless quarks near the transition temperature $T_c$.  The
corrections induced by this role to the parameters characterizing the 
fourth order terms in the Ginzburg-Landau free energy result in an increase 
of $\beta_1$ by a factor of $1+13C{\bar g}/2$ and an decrease of $\beta_2$ 
by a factor of $1-15C{\bar g}/2$.  The magnitude of these changes 
comes from the Meissner screening of the color magnetic force that dominates 
the interaction between quarks in the weak coupling regime as an essentially 
long-range force.  We thus see a contrast with the case of a short-range 
pairing interaction in which the corrections are suppressed by one power 
$T_c/\mu$.  We also note that the changes in $\beta_1$ and $\beta_2$ 
are in the direction of increasing the gap magnitude of the 2SC state and 
decreasing that of the CFL state.  This direction reflects the fact that 
not only the Meissner screening of the color magnetic force but also the 
color indices of magnetic gluons dominating the pairing interaction depend on 
the color structure of the pairing gap.  We have finally found that the 
feedback corrections, as long as ${\bar g}\ll 1$, keep the color-flavor 
locked phase the most stable just below $T_c$.

     The present result for the parameters $\beta_1$ and $\beta_2$, mainly 
through its effect on the gap magnitude near $T_c$, provides a way of 
studying strong coupling modifications on the previous weak coupling 
calculations based on the Ginzburg-Landau theory with $\beta_1=\beta_2$.  
Those calculations include the phase diagram \cite{I} as discussed in 
Sec.\ \ref{sec:phase} and its extension to nonzero quark masses \cite{IMTH}, 
responses to rotation and magnetic fields \cite{III,GR,cflmv}, and 
fluctuation-induced first order transition \cite{IV}.  Qualitatively, however,
no significant changes are expected.  Nonetheless, the tendency that 
$\beta_1>\beta_2$, if remaining at low densities, could be significant for 
the normal-super interfacial energy and the interaction between widely 
separated magnetic vortices for CFL quark matter.  This is because both 
quantities, which are sensitive to the ratio between $\beta_1$ and $\beta_2$ 
\cite{GR,cflmv}, control the criterion of whether or not the CFL state can 
allow magnetic vortices to form.

\acknowledgments

    We are grateful to Gordon Baym, Tetsuo Hatsuda, Taeko Matsuura, and Motoi 
Tachibana for helpful discussions.  We acknowledge the hospitality of the 
Institute for Nuclear Theory at the University of Washington, where this work 
was initiated.

\end{document}